\documentstyle[11pt,newpasp,twoside,graphicx,psfig]{article}
\def\edcomment#1{\iffalse\marginpar{\raggedright\sl#1\/}\else\relax\fi}
\reversemarginpar

\begin{document}
\title{Young star clusters: Clues to galaxy formation and evolution}
 \author{Peter Anders, Uta Fritze - v. Alvensleben}
\affil{Universit\"ats-Sternwarte G\"ottingen, Geismarlandstrasse 11, 37083 G\"ottingen, Germany}
\author{Richard de Grijs}
\affil{Department of Physics \& Astronomy, The University of Sheffield, Hicks Building, Hounsfield Road, Sheffield S3 7RH, UK}

\begin{abstract}
Young clusters are observed to form in a variety of interacting galaxies and violent starbursts, a substantial number resembling the progenitors of the well-studied globular clusters in mass and size. By studying young clusters in merger remnants and peculiar galaxies, we can therefore learn about the violent star formation history of these galaxies.
We present a new set of evolutionary synthesis models of our GALEV code specifically developed to include the gaseous emission of presently forming star clusters, and a new tool that allows to determine individual cluster metallicities, ages, extinction values and masses from a comparison of a large grid of model {\bf S}pectral {\bf E}nergy {\bf D}istributions (SEDs) with multi-color observations. First results for the newly-born clusters in NGC 1569 are presented.
\end{abstract}

\section{Models \& Applications}
We have further refined the G\"ottingen evolutionary synthesis code GALEV by including the effect of gaseous line and continuum emission. The emission contributes significantly to the integrated light (both to absolute magnitudes and colors) of stellar populations younger than $3 \times 10^7$ yr (see Fig. 1 and Anders \& Fritze - v. Alvensleben 2003). The updated models are available from
{\bf http://www.uni-sw.gwdg.de/$\sim$galev/panders/}.

\begin{figure}
\begin{center}
\includegraphics[angle=-90,width=0.64\linewidth]{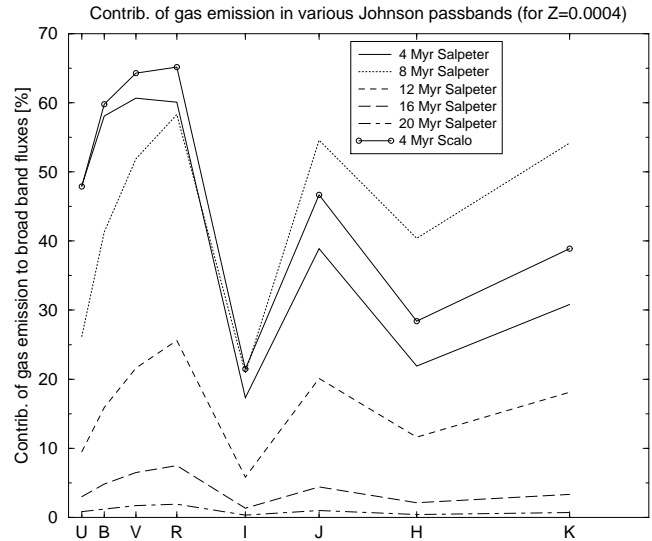}
\end{center}
\vspace{-0.3cm}
\caption{Contribution of gaseous emission to various standard Johnson magnitudes for metallicity Z = 0.0004, assuming a Salpeter or Scalo IMF as indicated in the legend.}
\end{figure}

We developed a tool to compare our evolutionary synthesis models with observed cluster SEDs, to determine the basic cluster parameters age, metallicity, internal extinction and mass independently. This tool was tested extensively using artificial clusters (Anders et al. 2003a) and broad-band observations of star clusters in NGC 3310 (de Grijs et al. 2003a). From our tests we conclude that:\\

\begin{enumerate}
\item At least 4 passbands are necessary to determine the 3 free parameters age, metallicity and extinction, and the mass by scaling the SED, independently.
\item The most important passbands are the {\sl U} and {\sl B} bands; for systems older than roughly 1 Gyr the {\sl V} band is equally important.
\item NIR bands improve the results by constraining the metallicity efficiently.
\item A wavelength coverage as long as possible is desirable.
\item Large observational errors and/or wrong {\sl a priori} assumptions may lead to completely wrong results.
\end{enumerate}

As a first application the dwarf starburst galaxy NGC 1569 was chosen. Our results are consistent with values in the literature, but enlarge the sample of star clusters studied in this galaxy by a factor of 3. We find a surprising change in the cluster mass function with age, with clusters formed during the onset of the burst (approx. 25 Myr ago) being on average more massive than the most recently formed clusters (Anders et al. 2003b).

\section{References}
Anders, P., Fritze - v. Alvensleben, U. 2003, \aap, 401, 1063\\
Anders, P., Bissantz, N., Fritze - v. Alvensleben, U., de Grijs, R. 2003a,\\ \mnras, {\sl submitted}\\
Anders, P., de Grijs, R., Fritze - v. Alvensleben, U., Bissantz, N. 2003b,\\ \mnras, {\sl submitted}\\
de Grijs, R., Fritze - v. Alvensleben, U., Anders, P., Gallagher, J. S., Bastian, N., Taylor, V. A., Windhorst, R. A. 2003a, \mnras, 342, 259\\
\end{document}